\documentclass[journal,12pt,onecolumn,draftclsnofoot,]{IEEEtran}
\usepackage{cite}

\ifCLASSINFOpdf
\else
\fi
\usepackage{bbm}

\usepackage{amsmath}
\usepackage{amssymb}

\usepackage{graphicx}
\usepackage{caption}
\usepackage{subcaption}

\usepackage[usenames, dvipsnames]{color}

\begin{document}
%
\title{Beam Based Stochastic Model of the Coverage Probability in 5G Millimeter Wave Systems}

\author{\IEEEauthorblockN{Cristian Tatino\IEEEauthorrefmark{1}\IEEEauthorrefmark{2}, Ilaria Malanchini\IEEEauthorrefmark{2}, Danish Aziz\IEEEauthorrefmark{2}, Di Yuan\IEEEauthorrefmark{1}}
\IEEEauthorblockA{\IEEEauthorrefmark{1}Department of Science and Technology, Link\"{o}ping University, Sweden\\
                Email: \{cristian.tatino, di.yuan\}@liu.se}      
                \IEEEauthorblockA{\IEEEauthorrefmark{2}Nokia Bell Labs, Stuttgart, Germany\\
                Email: \{ilaria.malanchini, danish.aziz\}@nokia-bell-labs.com }
}
\maketitle

\begin{abstract}
Communications using frequency bands in the millimeter-wave range can play a key role in future generations of mobile networks. By allowing large bandwidth allocations, high carrier frequencies will provide high data rates to support the ever-growing capacity demand. 
The prevailing challenge at high frequencies is the mitigation of large path loss and link blockage effects. Highly directional beams are expected to overcome this challenge. 
In this paper, we propose a stochastic model for characterizing beam coverage probability. The model takes into account both line-of-sight and first-order non-line-of-sight reflections. We  model the scattering environment as a stochastic process and we derive an analytical expression of the coverage probability for any given beam. The results derived are validated numerically and compared with simulations to assess the accuracy of the model.
\end{abstract}


%
\IEEEpeerreviewmaketitle

\section{Introduction}
The ever-growing data rate demand as well as the shortage of mobile frequency resources pose challenges for the upcoming fifth generation (5G) of mobile communications. A way to overcome these problems is to exploit unused frequency bands such as millimeter waves (mm-waves)  between 30 to 300 GHz. Mm-waves bring new opportunities, but at the same time raise challenges, e.g., the large path loss caused by higher frequencies dramatically reduces the cell coverage area \cite{RappPot}. The use of highly directional narrow beams with high beamforming gain can help in increasing the cell coverage distance \cite{mm-wave/Beam}, but it requires robustness in procedures such as initial access, beam tracking, mobility management, and handovers.

The main focus of ongoing research related to mm-wave communications is the study of propagation characteristics, channel modeling, beam forming, and medium access control design. Extensive research is still needed to enable mm-wave communications to be deployed in cellular systems. To this end, we provide a beam based stochastic model for evaluating the coverage probability for any given beam. The analytical expression derived can be then exploited for supporting system level optimization, such as mobility management.

\subsection{Related Works}
\label{sec:Contributions}
Communications using mm-waves have been initially investigated for indoor and short range applications, where propagation is facilitated by line-of-sight (LOS) conditions and low-mobility. 
In~\cite{BeamCodebook}, the authors propose two algorithms for beam searching, selection and tracking in wireless local area networks. They discretize the set of beams and find, by using iterative search, the best beam pair for the transmitter and the receiver. 
Similarly, the authors in~\cite{mmWaveBlockage3} develop a method that compensates link blockage by switching between the LOS link and a non line-of-sight (NLOS) link, whenever the former is blocked.
 However, they do not provide any analytical model of the beam coverage and blockage probability.

Lately,  the focus has shifted towards 
the application of mm-waves in outdoor scenarios and cellular systems. 
In~\cite{mmWaveRappport1, mmWaveRapp5}, the propagation characteristics of mm-waves are investigated. The study in \cite{mmWaveRappport1} collects measurements taken in New York at 28 and 38 GHz. Results show that, when a high directional antenna array is used, path loss does not create a significant impediment to the propagation and it is still possible to reach the typical cell coverage of a high density urban environment. 
Based on the measurements reported in~\cite{mmWaveRappport1},~\cite{mmWaveRapp5} derives a statistical channel model for the path loss, the number of spatial clusters, the angular dispersion, and the outage probability.

Other works exploit stochastic geometry in order to derive statistical channel models and analyze the performance of mm-wave cellular systems.
In~\cite{Tbai1}, by proposing a stochastic model for the scattering environment, the authors compute the transmitter-receiver link blockage probability and the probability of coverage both for low frequency and mm-wave cellular networks. However, reflections are ignored. 
In~\cite{Refl}, the authors propose an approach based on random shape theory to provide a statistical characterization of the mm-wave channel and to compute the power delay profile. The model takes into account both the LOS link and all the first-order reflections. Differently from our work, it considers omnidirectional antennas, at both the receiver and the transmitter, and it does not consider any beamforming approach.
Leveraging the results in~\cite{Tbai1}, a stochastic approach is adopted also in~\cite{Tbai2} to provide an analysis of the cell coverage probability and capacity.  
The authors demonstrate that, in high cell density conditions, mm-wave networks are able to provide sufficient signal-to-interference-plus-noise ratio (SINR) coverage and higher rate than the low-frequency cellular networks. Compared to previous works, they incorporate a directional beamforming for the SINR computation, at both the base station and the mobile user. However, in this case, reflections are ignored. The same assumption is done in \cite{Mac}, where the authors analyze the impact on the media access control layer design of highly directional communications for mm-waves. Using random shape theory, they investigate initial access and interference management, discussing handover and mobility issues.

\begin{figure*}[t!]
	\centering
	\begin{subfigure}{.27\textwidth}
		\centering
		\includegraphics[width=\linewidth]{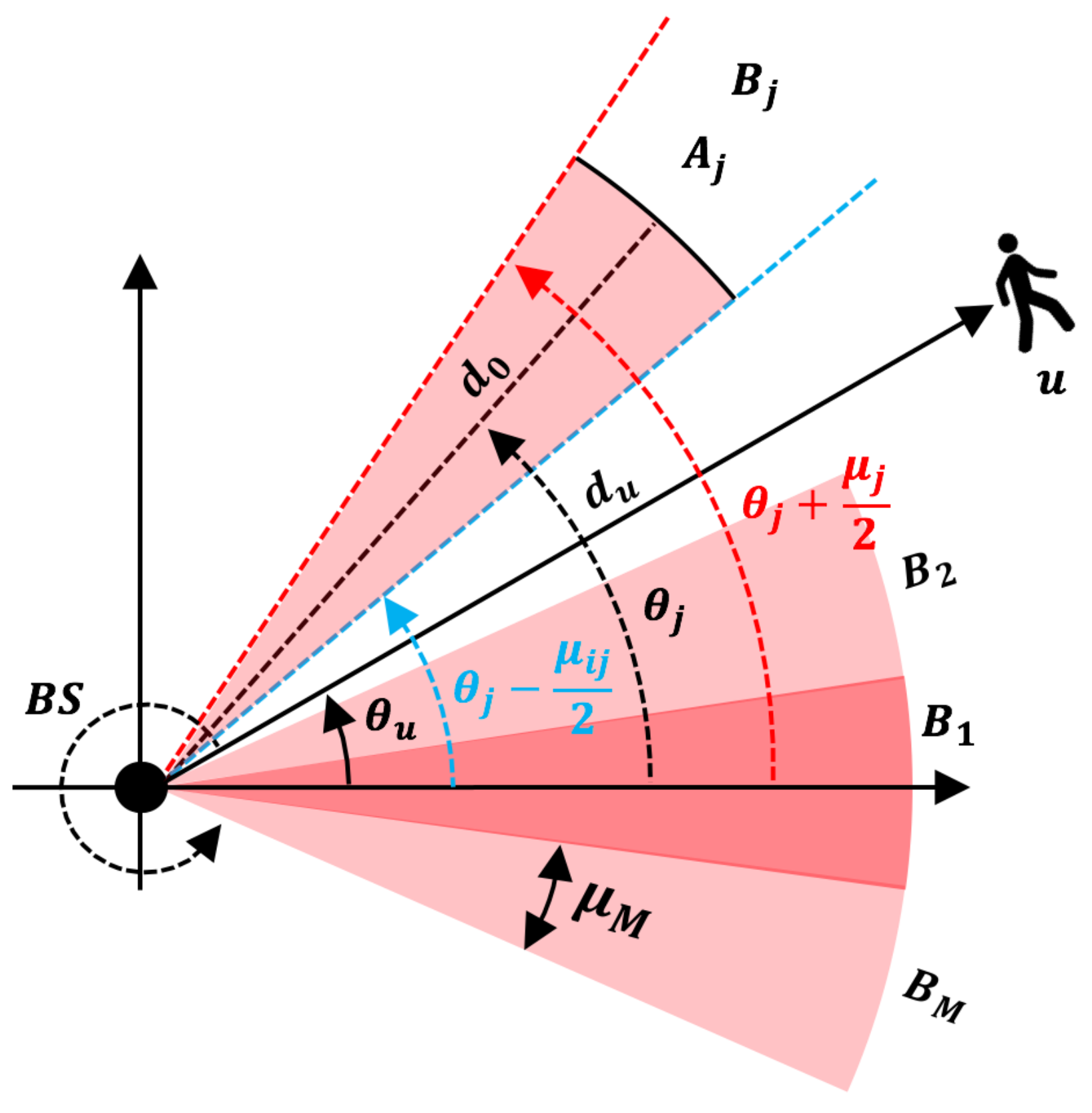}
		\caption{}
		\label{fig:Beams}
	\end{subfigure}%
	\begin{subfigure}{.38\textwidth}
		\centering
		\includegraphics[width=\linewidth]{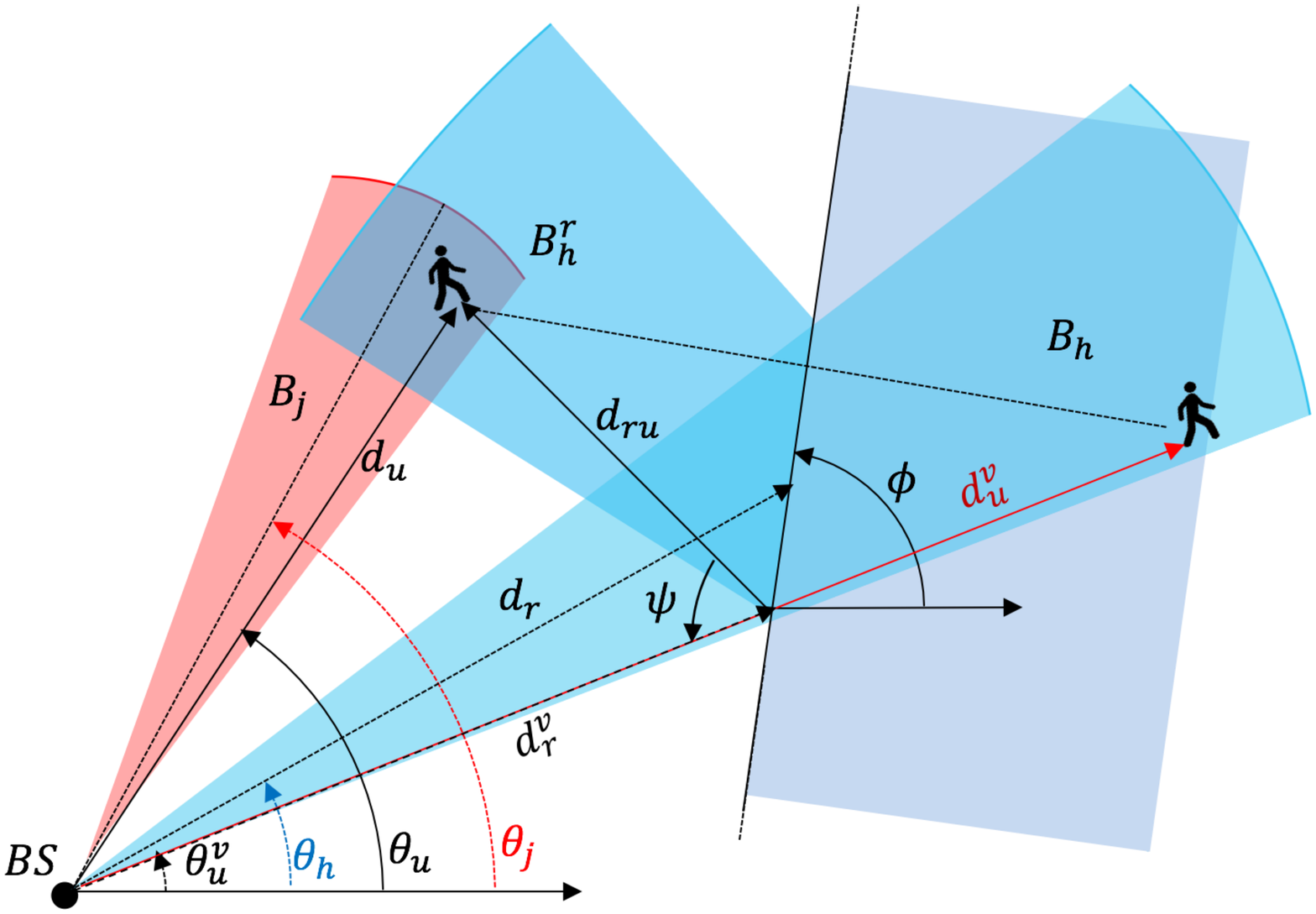}
		\caption{}
		\label{fig:Refl}
	\end{subfigure}
	\begin{subfigure}{.335\textwidth}
		\centering
		\includegraphics[width=\linewidth]{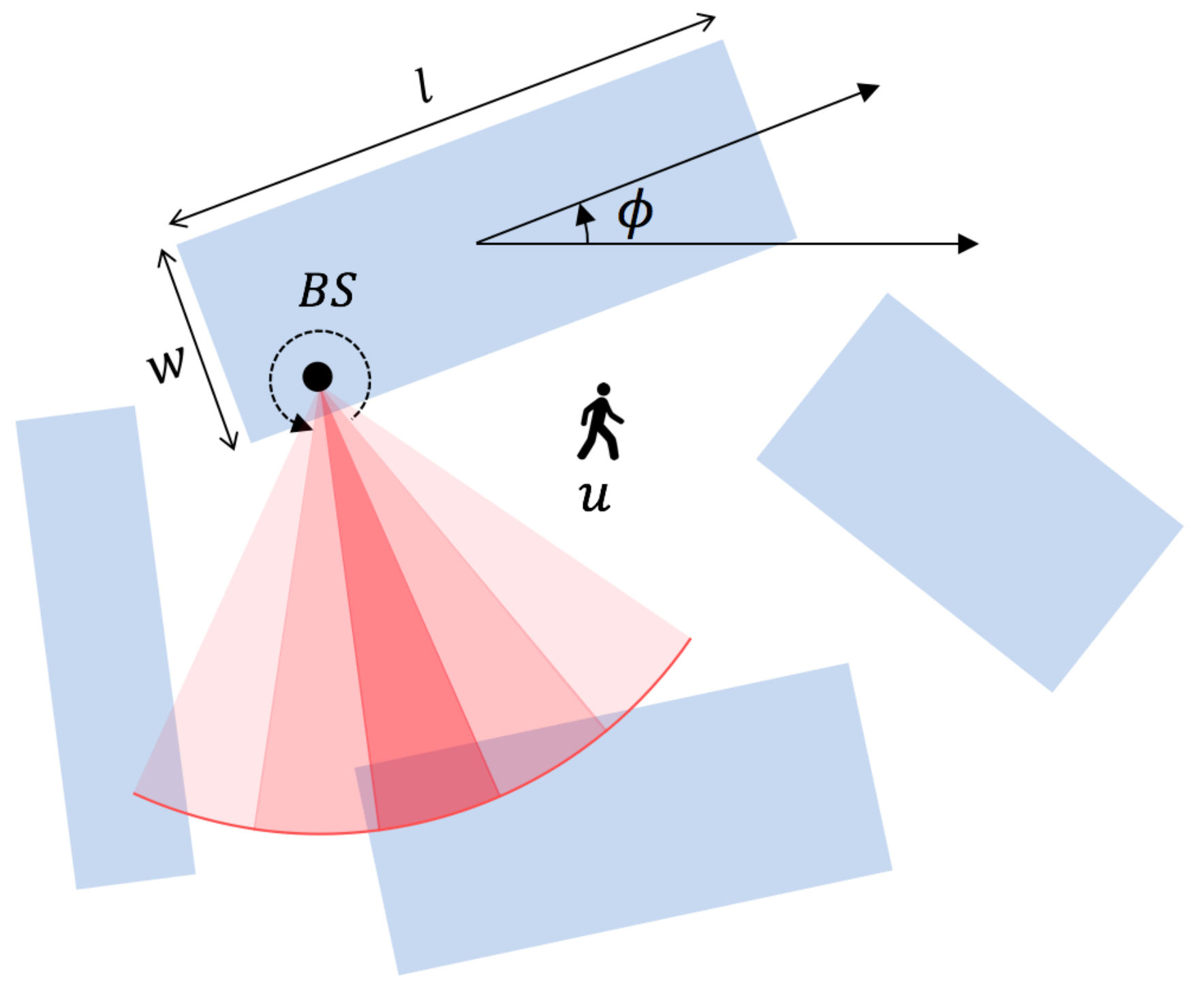}
		\caption{}
		\label{fig:General}
	\end{subfigure}
	\caption{(a) The BS and a subset of the beams that can be formed, indicated  with different color shades. (b) Beam $B_{j}$ covers directly the user $u$, whereas beam $B_{h}$ covers the user $u$ by the reflection $B_{h}^r$. (c) A mm-wave cell in an outdoor urban scenario with randomly deployed buildings.}
	\label{fig:General1}
\end{figure*}


\subsection{Our Contributions}
\label{sec:Contributions}
In this paper, we propose a beam based model that allows the analysis of the coverage probability and provides a useful tool to accurately investigate the effects of user mobility and beam selection in mm-wave cellular systems. 

The main novelty, with respect to the state of the art, lies in the coupling of the following two aspects.
\begin{itemize}
\item  The model incorporates beamforming, hence it allows to evaluate the coverage probability for any given beam, by taking into account the beam features (i.e., orientation and width) as well as the transmitter-receiver position.  
\item  The model evaluates the coverage probability not only considering the direct beam but also including first-order reflections, which fairly contribute to the coverage probability in NLOS conditions~\cite{mmWaveRappport1}.
\end{itemize}

The rest of the paper is structured as follows. Section \ref{sec:Ass} describes the system model and the assumptions. In Section~\ref{sec:Cov}, we derive the beam coverage probability when blockages and first-order reflections are taken into account. In Section~\ref{sec:Sim}, we present a numerical evaluation to validate the accuracy of the model proposed. Section~\ref{sec:Conclusion} concludes the paper.

\section{System Model and Assumptions}
\label{sec:Ass}


We target the analysis of beam coverage probability in a cellular scenario. For any cell and user position of interest, the analysis deals with the coverage probability of any given beam. To this end, we consider a cell using mm-wave frequency bands for radio access. 
The base station (BS) is in the center of the cell and is equipped with a linear array of antennas that can form a discrete set of beams $\mathcal{M}$ of cardinality $|\mathcal{M}|$. 
Each beam $B_{j}$, $j\in \mathcal{M}$, is defined using a sector model and is fully specified by its direction $\theta_{j}$ and width $\mu_{j}$, as shown in Fig.~\ref{fig:Beams}. Furthermore, we assume the beamforming gain $G_{j}$ to be a function of the beam width, i.e., $G_{j}=G(\mu_{j})$. 
The generic user \textit{u} is fully characterized by its position with respect to the BS, i.e., $(\theta_{u}, d_{u})$, which is given in polar coordinates as shown in Fig.~\ref{fig:Beams}, and is equipped with an omni-directional antenna. 

In order to compute the \textit{beam coverage probability}, we evaluate the signal-to-noise ratio (SNR) received by the generic user $u$ from the BS, when a certain beam is used to transmit. Namely, in our model, we assume that the SNR depends on the distance $d_{u}$, the carrier frequency $f$, and the blockage effects caused by the scattering environment. Moreover, we evaluate the SNR 
considering either the LOS link or a first-order reflection, while excluding the contributions given by links with two or more reflections. This is motivated by the fact that beams reflected more than once arrive at the receiver with a very high path loss (caused by the longer path and a larger reflection loss) and therefore we assume their contributions to the SNR to be negligible. In particular, a first-order reflection $B_{j}^r$ is generated when the beam $B_{j}$ hits a building, as shown in Fig.~\ref{fig:Refl}. We assume that beams are narrow enough to be totally reflected and we ignore diffraction and refraction effects. As a result, the beam $B_{j}$ can cover the user either directly or by its first-order reflection $B_{j}^r$. 
In order to model the scattering environment, as shown in Fig.~\ref{fig:General}, we consider buildings with rectangular shape. A building is specified by its center $z$, length~$l$, width $w$ and orientation $\phi$. We assume all these to be independent random variables. Namely, the centers of the buildings $Z$ form a homogeneous Poisson point process (PPP) of density $\lambda$. The lengths $L$ and the widths $W$ have probability density function  $f_{L}(l)$ and $f_{W}(w)$, respectively. The orientations $\Phi$ are assumed to be uniformly distributed between $[0, \pi]$.
A summary of the notation is reported in Table~\ref{glossary}.

\begin{table}[!t]
	\renewcommand{\arraystretch}{1.3}
	\caption{Summary of the notation}
	\label{glossary}
	\centering
	\begin{tabular}{r|l}
		\hline
		$B_{j}$ & $j^{th}$ beam\\
		$(\mu_{j},\theta_{j})$ & Width and orientation of $B_j$ \\
		$A_{j}$ & Sector covered by $B_j$\\
		$B_{j}^r$ & Reflected beam generated by $B_{j}$\\
		$(\theta_{u},d_{u})$ & Polar coordinates of $u$ \\
		$(\theta_{u}^{v},d_{u}^{v})$ & Polar coordinates of virtual user $u$ \\
		$d_{r}$ & Distance between the BS and the obstacle along $\theta_{h}$ \\
		$d_{r}^{v}$ & Distance between the BS and the obstacle along $\theta_{u}^{v}$\\
		$d_{ru}$ & Distance between the obstacle and $u$\\
		$\text{SNR}_{B_{j}u}$ & SNR for user $u$ and beam $B_{j}$ \\
		$\text{SNR}_{B_{j}u}^{D}$ & SNR for user $u$ and direct beam $B_{j}$ \\
		$\text{SNR}_{B_{j}u}^{R}$ & SNR for user $u$ and reflected beam $B_{j}^r$ \\
		$L$, $W$, $\Phi$ & Length, width and orientation of an obstacle \\
		\hline
	\end{tabular}
\end{table}

\section{Beam Coverage Probability}
\label{sec:Cov}
In this section, we compute the \textit{beam coverage probability} of beam $B_{j}$, by explicitly considering the dependency between this probability and the beam properties (i.e., orientation and width) as well as the position of the user. 
We define the event
$\mathsf{C}_{\mathsf{ju}}:=\text{SNR}_{{B_{j}u}} \ge \Gamma, $
where $\text{SNR}_{{B_{j}u}}$ is the SNR received by the user $u$ for beam $B_{j}$ and $\Gamma$ is a given threshold.
Formally, we define the \textit{coverage probability} of beam $B_{j}$ and the user~$u$, i.e., $P(\mathsf{C}_{\mathsf{ju}})$, as:
\begin{equation}\label{eq:prob_cov}
P(\mathsf{C}_{\mathsf{ju}})=P(\text{SNR}_{{B_{j}u}} \ge \Gamma).
\end{equation}

In order to compute the SNR received by the user at the position $(\theta_{u},d_{u})$, we distinguish between two cases: the beam $B_{j}$ covers the user directly or by a reflected beam $B_{j}^r$. Given the assumption that a beam is either not reflected or totally reflected by a building, we consider those two events to be mutually exclusive, i.e., the probability that the same beam covers simultaneously the user both directly and with a reflection is set equal to zero. 
A beam $B_{j}$ can cover the user directly if and only if the user $u$ is inside the sector $A_{j}$, which is defined by the direction $\theta_{j}$ and the width $\mu_{j}$ as:
\begin{equation}
\label{eq:Area}
A_{j}=\left\{(\theta,d) \in [0,2\pi]\times[0, \infty] : |\theta-\theta_{j}| \le \frac{\mu_{j}}{2} \right\}.
\end{equation}

Therefore, we define the event 
$ \mathsf{D}_{\mathsf{ju}}:=(\theta_{u}, d_{u}) \in A_{j}$
and we denote its complementary event as $\overline{\mathsf{D}}_{\mathsf{ju}}$.
Thus, according to the law of total probability, $P(\mathsf{C}_{\mathsf{ju}})$ can be written as:
\begin{multline}
\label{eq:coverage2}
P(\mathsf{C}_{\mathsf{ju}}) =P(\mathsf{C}_{\mathsf{ju}} \cap \mathsf{D}_{\mathsf{ju}})+P(\mathsf{C}_{\mathsf{ju}} \cap \overline{\mathsf{D}}_{\mathsf{ju}}) = \\
 P(\mathsf{C}_{\mathsf{ju}}|\mathsf{D}_{\mathsf{ju}}) P(\mathsf{D}_{\mathsf{ju}})+P(\mathsf{C}_{\mathsf{ju}} | \overline{\mathsf{D}}_{\mathsf{ju}}) P(\overline{\mathsf{D}}_{\mathsf{ju}}),
\end{multline}
where the first term of the sum represents the probability that the user is covered directly by the beam, while the second term is the probability to be covered by a reflection.


The first and the second addend of \eqref{eq:coverage2} are explicitly derived  in Section~\ref{sec:Direct} and in Section~\ref{sec:Refl}, respectively.

\subsection{Direct Beam Coverage Probability}
\label{sec:Direct}
The first term of \eqref{eq:coverage2} represents the probability of coverage with direct beam.
According to the definition of $\mathsf{D}_{\mathsf{ju}}$, we can write the probability  $P(\mathsf{D}_{\mathsf{ju}})$ as:
\begin{equation}
\begin{split}
\label{eq:Direct2}
P(\mathsf{D}_{\mathsf{ju}})= 
\begin{cases}
1 \hspace{2 cm} \forall~(\theta_{u},d_{u}) \in A_{j} \\
0 \hspace{2 cm} \mbox{otherwise.}
\end{cases}
\end{split}
\end{equation}

Note that the event $\mathsf{D}_{\mathsf{ju}}$ takes into account only whether the user lies in $A_{j}$ (or not). In order to incorporate the blockage effect of obstacles, we define $\text{LOS}_{u}$ ($\text{NLOS}_{u}$)  as the event in which the user $u$ is in LOS (NLOS) with respect to the BS. 
To compute the probability of $\text{LOS}_{u}$, we use one of the results derived in~\cite{Tbai1}. Namely, the authors show that (for the very same scattering model adopted here) the number of obstacles between the BS and the user is a random variable $O$ that follows a Poisson distribution with mean:
\begin{equation}
\label{eq:poiss}
E[O]=\beta d_{u}+p, \\
\end{equation}
where $ \beta= [ 2\lambda(E[L]+E[W]) ] \mathbin{/} \pi, ~p= \lambda E[L]E[W] $
and $E[X]$ indicates the mean of the random variable $X$. Therefore, the probability that the user is in LOS can be written as follows:
\begin{equation}
\label{eq:LOS}
\begin{split}
P(\mbox{LOS}_{u})=P(O=0)=e^{-(\beta d_{u}+p)}.
\end{split}
\end{equation}

Note that the two events $\mathsf{D}_{\mathsf{ju}}$ and $\mbox{LOS}_{u}$ are independent since the former depends only on $\theta_{u}$ and  $\theta_{j}$ whereas the latter depends only on $d_{u}$. 
Furthermore, since $\text{LOS}_{u}$ and $\text{NLOS}_{u}$ are complementary events, the first term of \eqref{eq:coverage2} can be rewritten as follows: 
\begin{multline}
	\label{eq:PC_Direct2}
	P(\mathsf{C}_{\mathsf{ju}} \cap \mathsf{D}_{\mathsf{ju}}) = \\
	P(\mathsf{C}_{\mathsf{ju}} |\mathsf{D}_{\mathsf{ju}} \cap \mbox{LOS}_{u}) 
	P(\mathsf{D}_{\mathsf{ju}})P(\mbox{LOS}_{u})+\\
	P(\mathsf{C}_{\mathsf{ju}} |\mathsf{D}_{\mathsf{ju}} \cap \mbox{NLOS}_{u})
	P(\mathsf{D}_{\mathsf{ju}})\left(1-P(\mbox{LOS}_{u})\right).
\end{multline}
Moreover, by assumption, refraction is not considered in our model and a signal is completely reflected by an obstacle, hence $P(\mathsf{C}_{\mathsf{ju}} |\mathsf{D}_{\mathsf{ju}} \cap \mbox{NLOS}_{u})=0$.

Let $\{\text{SNR}_{B_{j}u}|(\mathsf{D}_{\mathsf{ju}} \cap \mbox{LOS}_{u})\}$ be the received SNR when user~$u$ is directly covered by beam $B_{j}$ in LOS, which we indicate for the rest of the paper as $\text{SNR}_{B_{j}u}^{D}$. By applying the Friis' law we can write:
\begin{equation}
\label{eq:SNR_LOS}
\begin{split}
\text{SNR}_{{B_{j}u}}^{D}=\frac{P_{t}G_{j}G_{u}c^{2}}{ (4\pi d_{u} f)^{2} P_{N}},
\end{split}
\end{equation}
where $P_{t}$ is the transmit power, $c$ is the speed of light, $G_{u}$ is the user beamforming gain, $f$ is the frequency and $P_{N}$ is the noise power.

To compute the coverage probability, we consider the case in which the SNR is greater than the given threshold $\Gamma$. 
Thus, let us define $d_{0}$ as the distance for which $\text{SNR}_{{B_{j}u}}^{D}= \Gamma $; with $d_{0}$ and $A_{j}$ defining the \textit{beam coverage area} as shown in Fig.~\ref{fig:Beams}. 
Let $\mathbbm{1}_{\mathcal{X}}(x)$ be the indicator function, i.e., $\mathbbm{1}_{\mathcal{X}}(x)=1 ~\forall x \in \mathcal{X}$. We can then write the \textit{direct beam coverage probability} as:
\begin{multline}
\label{eq:P_SNR_LOS2}
P(\mathsf{C}_{\mathsf{ju}} \cap \mathsf{D}_{\mathsf{ju}})= P(\text{SNR}_{{B_{j}u}}^{D} \ge \Gamma) 
 P(\mathsf{D}_{\mathsf{ju}})  P(\mbox{LOS}_{u})=\\
 \mathbbm{1}_{[\theta_{j}-\frac{\mu_{j}}{2}, \theta_{j}+\frac{\mu_{j}}{2}] \times [0, d_0] }(\theta_{u},d_{u}) e^{-(\beta d_{u}+p)}.
\end{multline}

\subsection{Reflected Beam Coverage Probability}
\label{sec:Refl}
We now investigate the probability of being covered by a first-order reflection. 
In general, $B_{j}$ can generate different reflections, which depend on the position and orientation of the building that is hit by the beam. 
In order to compute them, we assume the specular reflection law, i.e., the incident angle is assumed to be equal to the reflected one. 
Moreover, given the narrow beam assumption, we consider only the case in which the entire beam hits only one side of an obstacle (see Fig.~\ref{fig:Refl}). 
Furthermore, the side of the building that is hit by the beam generates a straight line that divides the space in two half-planes, as shown in Fig.~\ref{fig:Refl}.
Thus, we compute the symmetric point $(\theta_{u}^v, d_{u}^v)$ of the user position with respect to this line, which we call  \textit{virtual user position}. 

The user $u$ is covered by a first-order reflected beam if the two events $\mathsf{R}_{\mathsf{ju}}$ and $\text{LOS}^{R}_{u}$ jointly hold,  
where $\mathsf{R}_{\mathsf{ju}}:=\{\theta_{u}^v \in A_{j}\} \cap \{d_{u}^v \ge d_{r}^v\}$
and $\text{LOS}^{R}_{u}$  is the event in which the user~$u$ is in LOS with respect to the obstacle. Note that $d_{r}^v$ is the distance between the BS and the obstacle along the direction identified by $\theta_{u}^v$, as shown in Fig.~\ref{fig:Refl}. By considering that the events $\overline{\mathsf{D}}_{\mathsf{ju}}$, $\mathsf{R}_{\mathsf{ju}}$ and $\text{LOS}^{R}_{u}$ are independent, we can write
\begin{equation}
\label{eq:PC_Refl}
P(\mathsf{C}_{\mathsf{ju}} | \overline{\mathsf{D}}_{\mathsf{ju}})=P(\text{SNR}_{{B_{j}u}}^{R} \ge \Gamma)P(\mathsf{R}_{\mathsf{ju}})P(\mbox{LOS}^{R}_{u}),
\end{equation}
where $\text{SNR}_{B_{j}u}^{R}$ is the received SNR when beam $B_{j}$ is reflected once by an obstacle. 
By applying the Friis' formula, we obtain 
\begin{equation}
\label{eq:SNR_LOS_R}
\begin{aligned}
\text{SNR}_{{B_{j}u}}^{R}=\frac{ P_{t}G_{j}^rG_{u} c^{2}}{(4\pi d_{u}^v f)^{2}\sigma P_{N}},
\end{aligned}
\end{equation}
where the beamforming gain of the reflected beam is 
$G_{j}^r=G_{j}$ and $\sigma$ is the reflection loss. 
Thus, we can derive the distance for which $\text{SNR}_{{B_{j}u}}^{R}= \Gamma$ as $d_{0}^v=\frac{d_{0}}{\sigma}$. 

The events $\{\text{SNR}_{{B_{j}u}}^{R} \ge \Gamma\}$, $\mathsf{R}_{\mathsf{ju}}$, and $\mbox{LOS}^{R}_{u}$ and their respective probabilities depend on, e.g.,  
$d_{u}^v$, $\theta_{u}^v$, and $d_{ru}$ (which is the distance between the user and the obstacle). Those in turn depends on the beam properties and the user position, which are both given, and on the 
 distance of the first obstacle from the BS, $d_r$, and its orientation $\phi$, see Fig.~\ref{fig:Refl}. According to the stochastic model adopted for the scattering environment, those variables are described by probability density functions  $f_{D_{r}}(d_r)$ and $f_{\Phi}(\phi)$, respectively. The latter is assumed to be uniformly distributed between $[0, \pi]$, whereas 
\begin{multline}
\label{eq:PDF}
f_{D_{r}}(d_r)= \\ \delta(d_r)\left(1-e^{-p}\right)+\left(1-\delta(d_r)\right)\beta e^{-\left( \beta d_r+p \right)}U(d_r),
\end{multline}
where $\delta(r)$ is the Dirac delta function, i.e.,  $\delta(r)=1$ for $r=0$ and $0$ otherwise, and $U(r)$ is the Heaviside step function.
The details of the computation can be found in Appendix A. 

To derive the final expression of the reflected beam coverage probability, reported in~\eqref{eq:PC_Refl2}, we condition all terms of~\eqref{eq:PC_Refl} on $D_{r}$ and $\Phi$. The product of the first two (conditioned) terms of~\eqref{eq:PC_Refl}  leads to the indicator function in~\eqref{eq:PC_Refl2}, whereas the $P(\text{LOS}^{R}_{u}|D_r=d_r, \Phi=\phi)$ is shown in Appendix~B.

\begin{figure*}[!t]
\normalsize
\begin{equation}
\label{eq:PC_Refl2}
\begin{aligned}
P(\mathsf{C}_{\mathsf{ju}}\cap \overline{\mathsf{D}}_{\mathsf{ju}})=(1-P(\mathsf{D}_{\mathsf{ju}}))\times \int_{o}^{\pi} \int_{0}^{\infty}& 
 \mathbbm{1}_{[\theta_{j}-\frac{\mu_{j}}{2}, \theta_{j}+\frac{\mu_{j}}{2}] \times [d_{r}^v, \frac{d_0}{\sigma}] }(\theta_{u}^v(r,\alpha),d_{u}^v(r,\alpha))
P(\text{LOS}^{R}_{u}|r,\alpha)\\
&\left(\delta(r)\left(1-e^{-p}\right)+\left(1-\delta(r)\right)\beta e^{-\left( \beta r+p \right)}U(r)\right)\frac{1}{\pi}  \mathrm{\,d}r \mathrm{\,d}\alpha
\end{aligned}
\end{equation}
\hrulefill
\end{figure*}

\section{Numerical Evaluation}
\label{sec:Sim}
In this section, we present the results of our study on the \textit{beam coverage probability}. 
We assess the validity of our model by comparing the numerical results for $P(\mathsf{C}_{\mathsf{ju}})$, computed using the analytical model, with simulation results.
We used Matlab to compute numerically~\eqref{eq:PC_Refl2}, hence~\eqref{eq:coverage2}, as well as to obtain the simulations results.
It is important to note that in the simulation setup, we remove the assumption that a beam can hit only one side of an obstacle and we allow the beam to hit several obstacles (and sides), hence generating several reflections. Clearly, this makes the simulation environment more realistic, but also leads to some gap between the model and the simulation results, as shown later. 

\subsection{Simulation Setup}
\label{sec:SimEnv}
We consider a simulation area of $500 \times 500$ m\textsuperscript{2} and we place the base station in the centre of the area. 
We independently generate $10,000$ instances by dropping the buildings randomly, according to a PPP of density $\lambda$.
In order to obtain a comprehensive performance evaluation, hereafter, we vary several parameters, such as beam width and orientation, building density, and position of the user. The parameters that are fixed are: $P_{t}=30$~dBm (as the experiments in \cite{mmWaveRapp5}), $P_N=-85$~dBm, $f=30 $ GHz, and $\Gamma=1$, i.e., $0$~dB. $W$ and $L$ are characterized by uniform distribution between $[30,50]$ and $[40,60]$ (in meters), respectively. The reflection loss, which depends on several factors, e.g., angle of incidence on the obstacle, frequency, materials of the wall, is set to $\sigma=3 $ dB (as proposed in \cite{Refl}), which means that half of the power is lost when the beam hits a building. Moreover, the results consider two different beam widths: $\mu_{ij}=10^{\circ}$ and  $\mu_{ij}=30^{\circ}$. Since the gain depends on the beam width itself, we set $G(10^{\circ})=36$ dBi and $G(30^{\circ})=12$ dBi, which are  assumed constant inside $A_{j}$ and equal to $0$ elsewhere. Moreover, we assume that $G_{u}=1$ dBi.

\begin{figure}[!tb]
\centering
\includegraphics[width=1\columnwidth]{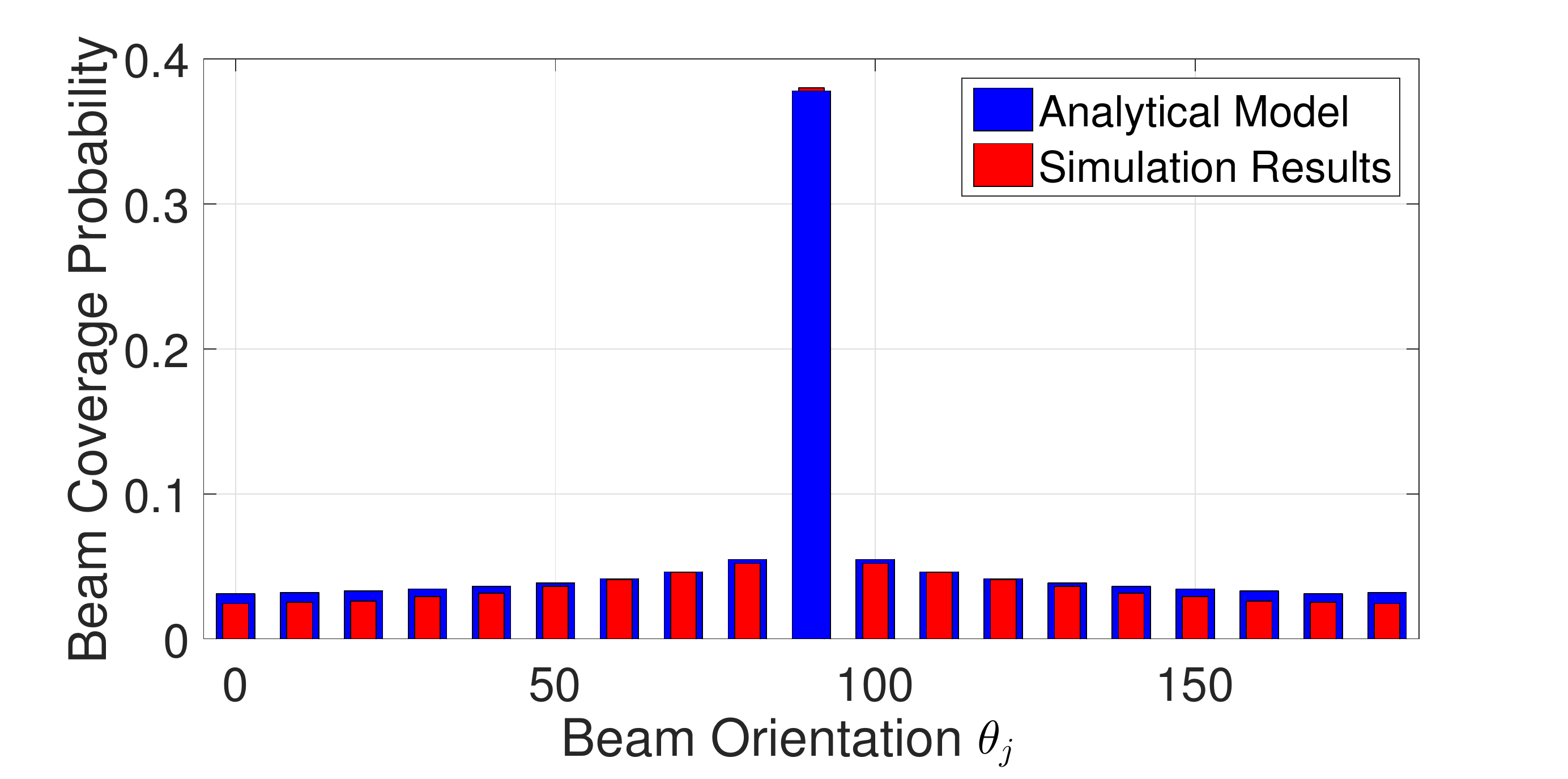}
\caption[]{The beam coverage probability computed for different non-overlapping beams with width $\mu_{j}=10^{\circ}$. }
\label{fig:Rosa}
\end{figure}

\subsection{Results}
\label{sec:SimEnv}
Fig.~\ref{fig:Rosa} shows the beam coverage probability when varying the beam orientation $\theta_{j}$. 
The user position is $(90^{\circ},50\mbox{ m})$, the building density is  $\lambda=0.0002$~buildings/m\textsuperscript{2}, and the beam width is $\mu_{j}=10^{\circ}$. 
First, we  observe that analytical and simulation results are very close to each other, validating the proposed model. Furthermore, we see that $P(\mathsf{C}_{\mathsf{ju}})$ decreases as the difference between the angular coordinate of the user, $\theta_{u}$, and the beam direction, $\theta_{j}$, increases. In particular, the coverage probability of the direct beam ($\theta_{j}=90^{\circ}$) is much larger than the ones obtained from reflected beams.  
Namely, when the beam direction moves away from the user angular coordinate, the path between the user and the obstacle becomes longer. Consequently, both the received SNR and the probability that the user is in LOS with respect to that obstacle decrease. 
Similar results have been obtained for different beam widths, but they are not reported for the sake of space. 

\begin{figure}[tb]
	\centering
	\includegraphics[width=1\columnwidth]{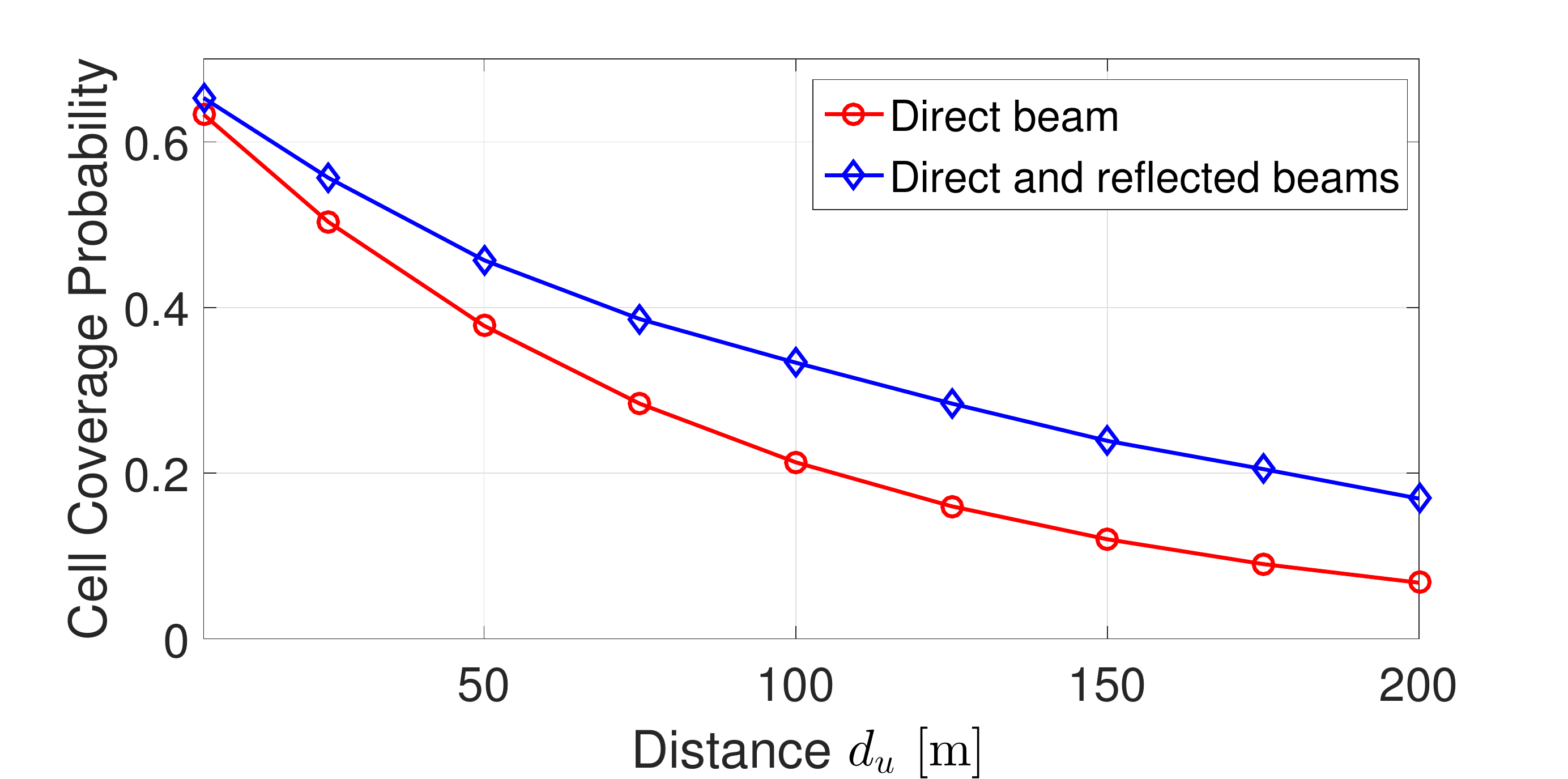}
	\caption[]{Comparison between the cell coverage probability when considering only the direct beam and when including all possible reflected beams. 
	}
	\label{fig:PCELL}
\end{figure}

\begin{figure}[tb]
	\centering
	\includegraphics[width=1\columnwidth]{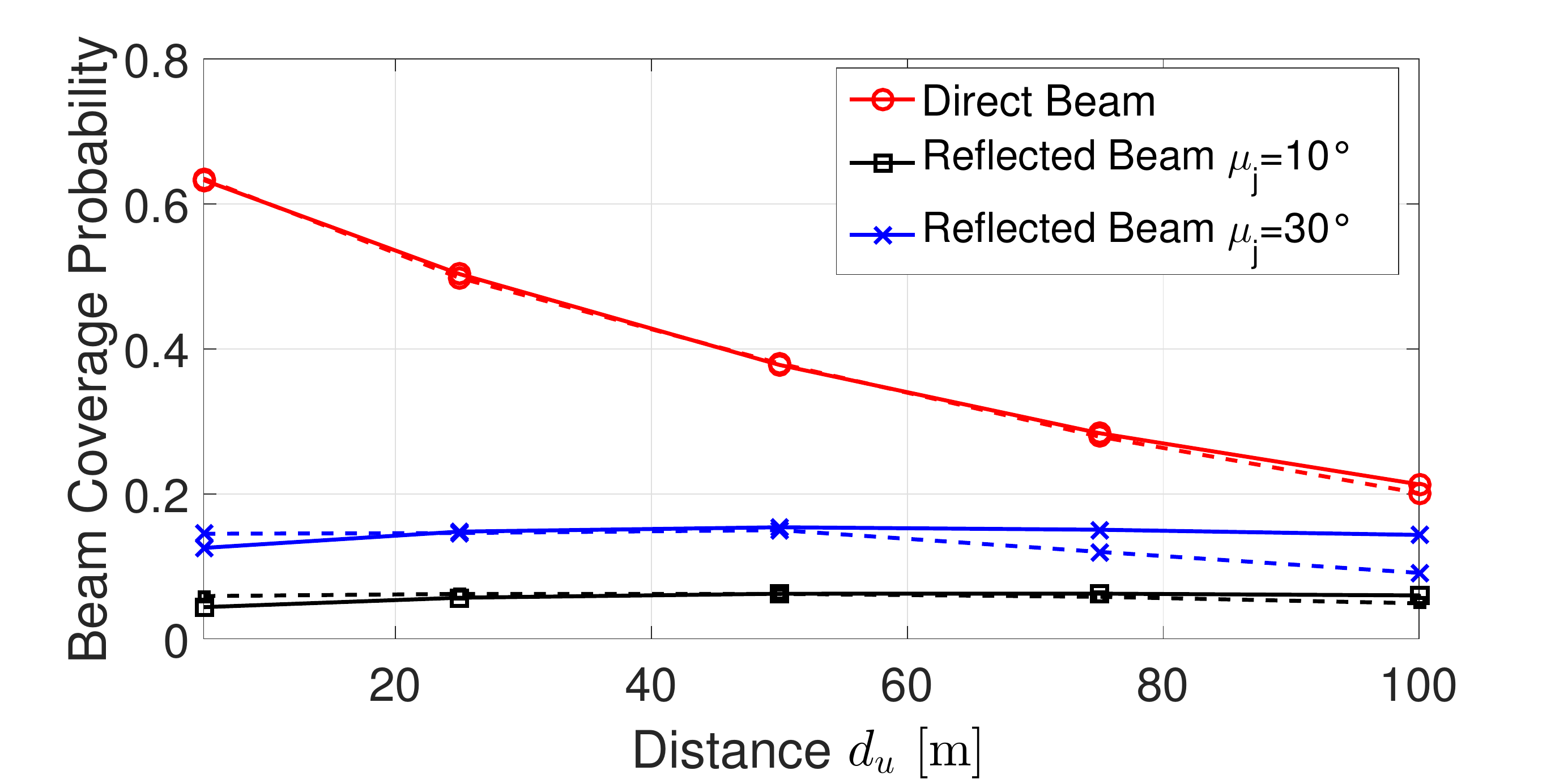}
	\caption[]{Comparison between the analytical model (solid line) and simulation results (dashed line) of the direct and reflected beam coverage probability with varying user distance~$d_{u}$. 
		\vspace{1.6mm}
	}
	\label{fig:Unico_dist}
\end{figure}

Although the contribution to the coverage probability of the non-direct beams is smaller compared to the direct one, the aggregation of all of them can have a significant impact on the cell coverage probability. 
In Fig.~\ref{fig:PCELL}, we compare the simulated cell coverage probability when only the direct beam is considered and when reflections are also included. Namely, in the latter case, the cell coverage probability is defined as the probability that at least one beam covers the user. We use the same parameters of Fig.~\ref{fig:Rosa}, but we vary the user distance~$d_u$. We observe that the cell coverage probability of the direct beam decreases faster  in comparison to that with reflected beams. Moreover, the contribution of the reflections become more evident as the user distance increases. 
Similar conclusions can be drawn from Fig.~\ref{fig:Unico_dist}, in which we show the beam coverage probability, varying the user distance $d_{u}$, for the direct beam and two reflected beams, when $\lambda=0.0002$~buildings/m\textsuperscript{2}. In this case, the angular coordinate of the user is $\theta_{u}=90^{\circ}$ and the direct beam is the one with orientation $\theta_{j}=\theta_{u}$. Furthermore, we select two reflected beams: a first beam with $\theta_{j}=95^{\circ}$ and $\mu_{j}=10^{\circ}$, and a second beam with $\theta_{j}=105^{\circ}$ and $\mu_{j}=30^{\circ}$. In both cases, the reflected beam is chosen such that the user is placed at the border of the beam coverage angle, which corresponds to the best non-direct beam (in terms of beam coverage probability). 
We observe that the direct beam coverage probability decreases rapidly as $d_{u}$ increases, whereas the reflected beam coverage probability remains almost constant. 
Fig.~\ref{fig:Unico_dist} further validates our model, reported with solid lines, with respect to simulations results, reported with dashed lines. In particular, the divergence between model and simulation results, for the reflected beam curves, increases with $d_u$. This is due to the assumption (made in the analytical model, but not in the simulation setup) that the reflected beam hits only one side of the same building and is totally reflected, proving that our model is more accurate for narrow beams. 

\begin{figure}[tb]
	\centering
	\includegraphics[width=1\columnwidth]{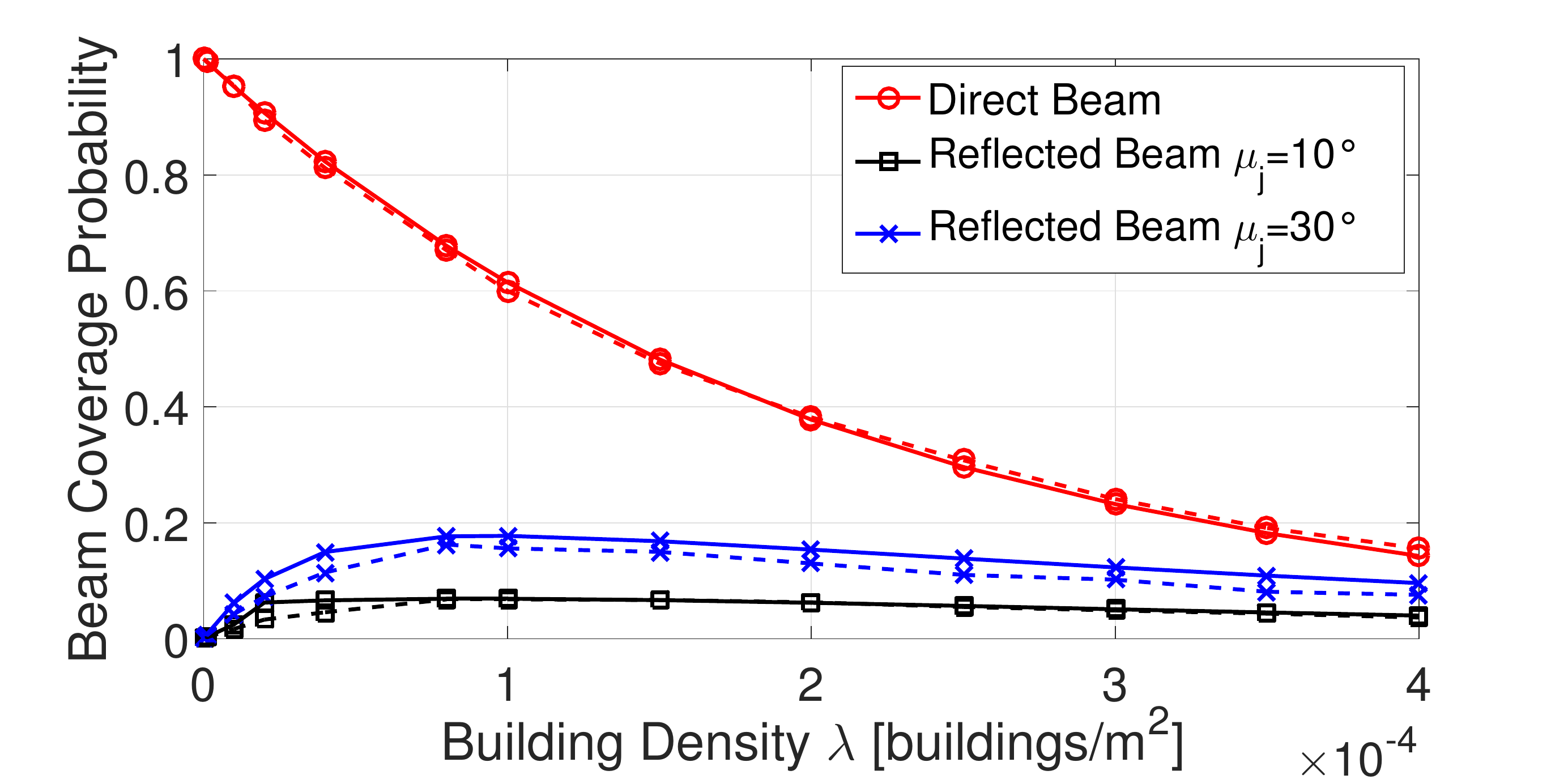}
	\caption[]{Comparison between the analytical model (solid line) and simulation results (dashed line) of the direct and reflected beam coverage probability with varying building density $\lambda$. 
	}
	\label{fig:Unico_lambda}
\end{figure}

The validity of the model is shown also in Fig.~\ref{fig:Unico_lambda}, where we compare the beam coverage probability when the building density $\lambda$ increases, for a fixed user position $(90^{\circ},50\mbox{ m})$ and the same beam set assumed in Fig.~\ref{fig:Unico_dist}. 
In general, we observe that the analytical model matches well the simulation results. Moreover, the coverage probability of the direct beam decreases as the density $\lambda$ increases, whereas the coverage probability of the reflected beams has a non-monotonic behavior. Namely, it increases from zero (when there are no buildings, hence no reflections) reaching a maximum for a given building density, and then decreases again.
This behaviour is due to twofold effect that the building density has on the reflections. On one hand, increasing the building density corresponds to increasing the possibilities of generating reflections, which enhances the beam coverage probability. On the other hand, increasing the building density reduces the probability of LOS between the position of the first obstacle and the user, which decreases the beam coverage probability. 

Finally, both Fig.~\ref{fig:Unico_dist} and Fig.~\ref{fig:Unico_lambda} show that, by increasing the beam width, we can enhance the coverage probability of reflected beams (i.e., when there are NLOS conditions). This is due to the fact that the reflection of wider beams
can cover a larger area and thus increase the probability of covering the user. This is an important outcome of our analysis, which suggests that  width should be trade off between narrow beams, which are very good in LOS conditions, and wider beams, which can provide good coverage probability in NLOS conditions.


\section{Conclusion}
\label{sec:Conclusion}
In this paper, we propose a beam based stochastic model for evaluating the beam coverage probability in mm-wave cellular systems. We model the scattering environment as a stochastic process and we derive an analytical expression valid for any given beam with respect to a given user position.  The proposed model is able to capture the dependency of the beam coverage probability on various  parameters, such as user position, beam orientation and width, and building density. 

In general, the analytical model matches well the simulation results, especially for narrow beams. Furthermore, we show that, although the highest coverage probability is provided by the direct beam, reflections can fairly contribute to it, especially for larger user distances, i.e., when the LOS probability dramatically decreases.  Moreover, the results show a non-monotonic behaviour of the reflected beam coverage probability with respect to the building density, which suggests that an optimal building density exists for NLOS conditions. Finally, we observe that increasing the beam width is a good strategy to improve the beam coverage probability in NLOS~conditions.

Future work will further investigate the coverage properties due to reflections, and extend the model to improve the accuracy for wider beams. Furthermore, we will investigate how the proposed model can be used for network optimization, e.g., mobility management,  in mm-wave systems.

\section{Appendix A}
\label{sec:distance}
Hereafter, we derive the probability density function (PDF) of the distance of the first obstacle from the base station along a given direction, i.e.,  $f_{D_{r}}(d_r)$.
Recall that the distribution of the total number of obstacles along a particular segment of distance $d_r$ is a Poisson random variable $O$ with mean defined in~\eqref{eq:poiss}, cf.~\cite{Tbai1}. Therefore, the  cumulative density function (CDF) $F_{D_{r}}(d_r)$ can be written as
\begin{multline}
\label{eq:CDF_D}
F_{D_{r}}(d_r) = P(D_{r} \le d_r)=1-P(D_{r} \ge d_r) = \\
1-P(O(d_r)=0)=1-e^{-(\beta d_r + p)}.
\end{multline}

Since the CDF is equal to 0 for $d_r<0$, it has a discontinuity in zero caused by the non-zero dimension of the obstacles. Therefore, to compute the PDF $f_{D_{r}}(d_r)$ we separate the two cases, i.e., $d_r=0$ and $d_r>0$. Then we obtain 
\begin{equation}
\begin{split}
\label{eq:PDF}
f_{D_{r}}(d_r)= 
\begin{cases}
1-P(O(0)=0)=1-e^{-p} &  d_r=0\\
\frac{\mathrm{\,d}F_{D_{r}}(d_r)}{\mathrm{\,d}d_r}=\beta e^{-\left( \beta d_r+p \right)} &  d_r>0.
\end{cases}
\end{split}
\end{equation}

\section{Appendix B}
\label{sec:PLOS}
One can easily see that the event $\text{LOS}^{R}_{u}$ is strongly correlated to the distance, $d_r$, between the BS and the first obstacle. 
For the sake of space we skip the details and we directly report the derived
 approximation of $P(\text{LOS}^{R}_{u}| D_r=d_r, \Phi=\phi)$, which is
\begin{multline}
\label{eq:PLOSr}
P(\text{LOS}^{R}_{u}|D_r=d_r, \Phi=\phi) = \\
\begin{cases}
\min(1,e^{-\beta d_{ru}+q}) & 0 \le \psi \le \frac{\pi}{2}, d_r\ge d_{ru}\\
\min(e^{-\beta (d_{ru}-d_r)},e^{-\beta d_{ru}+q}) & 0 \le \psi \le \frac{\pi}{2}, d_r\le d_{ru} \\
e^{-\beta d_{ru}} &  \frac{\pi}{2} \le \psi \le \pi
\end{cases}
\end{multline}
where $d_{ru}$ is the distance between the user and the obstacle, $q=\lambda \cot (\psi) (E[L^{2}]+E[W^{2}]) / 2$ and $\psi$ is the angle formed by the reflection and directly depends on $\phi$, as shown in Fig.~\ref{fig:Refl}. $E[L^{2}]$, and $E[W^{2}]$ are the second moments of the length and width of the obstacles, respectively. 

\section*{Acknowledgment}
The authors would like to thank Dr. Vangelis Angelakis and Dr. Nikolaos Pappas for the insightful discussions. 

This project has received funding from the European Union's Horizon 2020 research and innovation programme under the Marie Sklodowska-Curie grant agreement No. 643002.

\bibliographystyle{IEEEtran}
\bibliography{ref}

\end{document}